\documentclass[showpacs,aps,prb,amsmath,amssymb]{revtex4}
\begin{document}
\title{t-J model then and now: A personal perspective from the pioneering times}
\author{Jozef Spa\l{}ek}
\affiliation{Marian Smoluchowski Institute of Physics, Jagiellonian University, ul. 
Reymonta 4, 30-059 Krak\'ow, Poland}

\begin{abstract}
In this overview I sketch briefly the path to the so-called
{\em t-J model\/} 
derived for the first time 30 years ago
and provide its original meaning within the theory of strongly 
correlated magnetic metals with a non-Fermi (non-Landau) liquid ground state.
An emergence of the concept of {\em real space pairing\/}, 
is discussed in a historical prospective. A generalization
of this model to the many-orbital situation is briefly discussed.
The emphasis 
is put on didactical exposition of ideas, as they were transformed into
mathematical language.
The concept of {\em hybrid pairing\/} is introduced in the same
context at the end.
\vspace{2mm}
\end{abstract}

\maketitle

\section{Introduction}

Before presenting the details, let me first summarize the principal features of 
the so-called {\em t-J model\/}. This model represents a very nontrivial model of 
{\em strongly correlated fermions\/} because of the following principal reasons:

\begin{enumerate}

\item It represents a {\em system\/} of strongly inter-correlated itinerant 
fermions which transform into an antiferromagnetic state of localized particles
(the {\em Mott-Hubbard insulator\/}); the classic situation takes place at the
concentration of 
one fermion per single-band state (at the half filling, $n=1$).
The half-filled state is an antiferromagnetic insulator modeled by the
{\em Anderson kinetic exchange\/}.

\item The itinerant state is represented by particles called the {\em 
correlated holes\/} or simply the {\em holes\/} in the Mott-Hubbard insulator, 
which do not have the 
ordinary fermion properties (their creation and annihilation operators do
not obey the fermionic anticommutation rules). In other words, they cannot be
represented by Landau quasiparticles in an {\em exact\/} manner, 
since they do not represent almost-filled band states.

\item Magnetic interaction between the correlated itinerant
particles, again {\em the 
kinetic exchange\/}, is regarded also as the source of {\em real-space 
pairing\/} as it is represented as
taking place between the nearest neighbors. Hence, antiferromagnetism
and this new type of paired state must be regarded on {\em equal footing\/}
with the paired state (resonance-valence bond state or 
superconducting state).
This is particularly relevant in the context of high-temperature
superconductivity near the band filling, i.e. close to the Mott transition.

\item Because of the nontrivial character of the kinetic- 
(or residual-band-) energy
({\em the projected hopping part\/}), it can become comparable or even smaller
than the kinetic-exchange-energy part. In effect, magnetic polaron or 
phase-separated states or new type of spin-paired states can be formed,
concomitantly with the transition to a localized state for small carrier 
concentration of holes. Thus the transition in real systems such as
La$_{2-x}$Sr$_{x}$CuO$_{4}$, YBa$_{2}$Cu$_{3}$O$_{6+x}$, 
La$_{1-x}$Sr$_{x}$TiO$_{3}$, takes place even for a non-half-filled
band configuration. Although, here the role of atomic disorder is 
probably also very important, if not crucial.

\end{enumerate}

The four mentioned above features originate from the circumstance, that the
kinetic (band) energy of the fermionic particles in the correlated systems
is relatively small and easily comparable (if not smaller) to the 
Coulomb interaction 
energy. Furthermore, both energies are then 
effectively counted on the Kelvin, rather
than on the electronovolt scale and represent competing compensating
each other, dynamical contributions to the electron states. 
Hence, the system is very susceptible
to the perturbations such as thermal or atomic disorder, or the
electron-lattice coupling. This sensitivity of the 
strongly correlated metallic (or {\em magnetic-
insulating\/}) state leads to an instability with respect to the perturbations 
on the thermal 
scale and complicates enormously a reliable solution of the model, in addition 
to the 
specific situation that, unlike in the standard quantum mechanics, there is 
no small parameter in the model, since the potential (interaction) energy
is at least comparable to the single-particle (band) energy. 

The structure of the paper is as follows. In Section 2 we define the limit
of strong correlation. In Section 3 we discuss the original derivation
and the meaning of t-J model. In Section 4 we rewrite the exchange part
in terms of pairing operators and discuss this new aspect briefly,
whereas in Section 5 we introduce hybrid pairing.

\section{The Hubbard model}

The Hubbard model (1963) [1], 
was devised first to describe the single-band 
magnetism, particularly
to understand the so-called {\em itinerant\/} (Stoner-Wolfharth) ferromagnetism
and criterion of its appearance in a microscopic manner. 
It is represented by the Hamiltonian, which 
in the {\em second-quantization representation\/} has the form
\begin{equation}
H\,=\,\sum_{ij\sigma} {'}\, t_{ij}\, a_{i\sigma}^{\dagger}\, a_{j\sigma}\,+\,
U\,\sum_{i}\,n_{i\uparrow}\, n_{i\downarrow}\,,
\label{eq:w1}
\end{equation}
where the primed summation means that we take only $i\neq j$ terms in the 
{\em hopping\/} part.
The ingenuity of this expression derives from the fact, that we start from the 
atomic (Wannier) representation rather than from the popular then electron-gas
(band) representation of the many-electron state in solid. This representation 
(and actually, the Hamiltonian), has been invented earlier [2], although the 
meaning of the model was in the latter case quite different namely, to derive 
the
antiferromagnetic superexchange interaction as an effective $d-d$ exchange
in the Mott insulators (such as NiO, MnO or CoO),
without involving, explicitly the filled $p$ shells of oxygen
O$^{2-}$ or other chalcogenide ions. 

Obviously, one should think that the Bloch $\{\Phi_{{\bf k}}({\bf r})\}$ and the 
Wannier $\{W_{i}({\bf r})\}$ 
single-particle bases are physically equivalent, as they are 
related by (unitary) Fourier transform on the lattice

\begin{equation}
\left\{
\begin{array}{c}
\Phi_{{\bf k}}({\bf r})\,=\, \frac{1}{\sqrt{N}}\,\sum_{i}\, W_{i}({\bf r})\,
e^{i\, {\bf k}\,\cdot\,{\bf R}_{i}} \,, \\
W_{i}({\bf r})\,=\,\frac{1}{\sqrt{N}}\,\sum_{{\bf k}}\,
\Phi_{{\bf k}}({\bf r})\,
e^{-i\, {\bf k}\,\cdot\,{\bf R}_{i}}\,.
\end{array}
\right.
\label{eq:w2}
\end{equation}

Likewise, the same unitary equivalence holds between the annihilation operators
in the Bloch $\{ a_{{\bf k}\sigma}\}$ and the Wannier $\{ a_{i\sigma}\}$ namely

\begin{equation}
\left\{
\begin{array}{c}
a_{{\bf k}\sigma}\,=\, \frac{1}{\sqrt{N}}\,\sum_{i}\, a_{i\sigma}\,
e^{i\, {\bf r}\,\cdot\,{\bf R}_{i}}\,, \\
a_{i\sigma}\,=\,\frac{1}{\sqrt{N}}\,\sum_{{\bf k}}\, a_{{\bf k}\sigma}\,
e^{-i\, {\bf k}\,\cdot\,{\bf R}_{i}}\,.
\end{array}
\right.
\label{eq:w3}
\end{equation}

The main (and fundamental) aspect of this problem is that, depending on the
value of 
$U$, the equivalence between the two representations can be broken and, as $U$ 
increases, the atomic representation of electronic states 
becomes more appropriate than the 
Bloch one.
This fact expresses the so-called {\em Mott-Hubbard\/} or
{\em metal-insulator transition\/}, which in this case, is associated with the
breakdown of the global $U(1)$ symmetry, as in the Mott-Hubbard
insulating phase linear momentum is in an obvious manner not a proper quantum
number. Parenthetically, one should say that the Hubbard (1964) method of 
defining the localization [3] is similar to that of Mott. 
In essence, the Hubbard approach takes into account both the 
existence of crystal lattice, as well as the narrow-band structure of the 
correlated electrons. Explicitly, taking expectation values of the two 
terms in (\ref{eq:w1}) one sees, that in the Hartree-Fock approximation
and for one electron per atom we have the ground state energy per site
in the form

\begin{equation}
\frac{E_{g}}{N}\,\simeq\, -\vert \,\sum_{j(i)}\,t_{ij}\,\vert\,
\sum_{\sigma}\,n_{\sigma}\, (1\,-\, n_{\sigma})\,+\,
U\, n_{\sigma}\, n_{\overline{\sigma}}\,
%\stackrel{\rm n=1}{=}\,
\,=\, -\frac{W}{4}\,+\,\frac{U}{4}\,,
\label{eq:w4}
\end{equation}
where $W$ is the bandwidth and $z$ number of nearest neighbors
($W=2z|t|$). In the paramagnetic state ($n_{\sigma}= n_{\overline{\sigma}}=
1/2$) therefore, the two energies compensate each others when $U=W$ i.e.
when band and the Coulomb energy are comparable. The true value $U=U_{C}$
of critical interaction depends on the method selected and the density of
states for given system, but it is in the regime 
$W\lesssim U_{C}\lesssim 2W$ [4]. In the limit $U<<W$ we have the so-called
{\em metallic limit\/}. The question interesting us here is the {\em limit
of strong correlations $U>>W$\/}. In the case of $n=1$ this limit corresponds
to the Mott insulating limit [2,4,5]. The question which was posed by us, I 
believe
for the first time [6-7], was what happens when $U>>W$ and $n\neq 1$.

The situation for $n\neq 1$ (it is sufficient to consider the case $n<1$ only,
as that with $n>1$ is related to the former by switching to the hole language).
Namely, the Hartree-Fock approximation for $U\simeq 2W$ we have from 
(\ref{eq:w4})

\begin{equation}
\frac{E_{g}}{N}\,=\, -\frac{1}{2}\,
W\, n\, \left( 1\,-\,\frac{n}{2}\right)\,+\,
U\,\frac{n^{2}}{4}
\label{eq:w5}
\end{equation}

Taking $n=1-\delta$, with $\delta <<1$, we have that $E_{g}\sim W\delta/4$
i.e. the energy is positive and thus the metallic state is unstable. We show
below that this is indeed the case for $\delta<\delta_{C}\sim 0.1$, but
that situation requires 
a refined analysis. In any case, if we take Eq. (\ref{eq:w5}) literally
then for every $n$ we can define $U_{C}$ at which the two energies cancel out
each other
i.e. $U_{C}/W=(2-n)/n$, ($U_{C}$ increases as $n$ decreases). This
trivial reasoning teaches us one important thing: the creation of holes in the
Mott insulator makes particles effectively less correlated for fixed value of 
$U$. In other words the condition $U>>W$ valid for a half-filled band system
gets softened when we change (diminish) the number of electrons in the same 
band. In our derivation of the t-J model [6,7] and in subsequent papers it is 
assumed that the particles are strongly correlated in the whole filling range
considered. Note however that if we take $n=0.7$ then $U_{C}/W=1.86$, i.e.
the critical value is almost doubled, so if we had band states with the value of 
$U$ 
such that $U/U_{C}=1$ for $n=1$, for $n=0.7$ the particles behave as if they 
experienced only half of the critical value (are {\em substantially\/}
less correlated). This must be remembered particularly in the context of
high-temperature superconductors described via a single-band model since
then for $n=1$ we have $U/W\sim 2\div 3$ [8].

We provide next our original version of the derivation of t-J model [6,7]
and dwell on the details, which seemed to us not obvious whatsoever and are 
usually overlooked in the textbooks or review articles.

\section{t-J model as it was derived then}

I became interested in the Hubbard model around 1974 an particularly, in the 
problem of deriving its ferromagnetic state in the limit $U>>W$. This was 
because on the one-hand we have a definite result of Anderson from 1959 [2] 
that the Mott insulators are antiferromagnetic and from the other that of
Nagaoka [9] that the ground state of the Mott insulator in the limit $U=\infty$
is for certain lattices (e.g. fcc) ferromagnetic. The obvious thing was to
combine the two. In that manner, the question how to generalize the Anderson 
kinetic-exchange Hamiltonian to the metallic regime has arisen naturally, but 
the 
question was how to do it in detail. As the kinetic exchange is $\sim t^{2}/U$,
one should get it from the second-order virtual hopping processes, with the 
doubly
occupied site configurations in the intermediate state. For that purpose, one 
can rewrite the dynamical processes contained in the hopping term as follows

%\begin{equation}
$$
a_{i\sigma}^{\dagger}\, a_{j\sigma}\,\equiv\, a_{i\sigma}^{\dagger}\,
(1\,-\, n_{i\overline{\sigma}}\,+\, n_{i\overline{\sigma}})\, a_{j\sigma}
\,(1\,-\, n_{j\overline{\sigma}}\,+\, n_{j\overline{\sigma}})\,=\,
a_{i\sigma}^{\dagger}\,(1\,-\, n_{i\overline{\sigma}})\, a_{j\sigma}\,
(1\,-\, n_{j\overline{\sigma}})
$$
\begin{equation}
+\,a_{i\sigma}^{\dagger}\,
(1\,-\, n_{i\overline{\sigma}})\, a_{j\sigma}\, n_{j\overline{\sigma}}\,+\,
a_{i\sigma}^{\dagger}\, n_{i\overline{\sigma}}\, a_{j\sigma}\, 
(1\,-\, n_{j\overline{\sigma}})\,+\,
a_{i\sigma}^{\dagger}\, n_{i\overline{\sigma}}\, a_{j\sigma}\, 
n_{j\overline{\sigma}}\,.
\label{eq:w6}
\end{equation}

The consecutive terms represent the four restricted types of hopping processes:
the first hopping from singly occupied site onto an empty one, the second 
hopping from singly occupied site and formation of a doubly occupied one, etc.
In such manner, the full hopping term looks like

%\begin{equation}
$$
\sum_{ij\sigma} {'}\, t_{ij}\,
a_{i\sigma}^{\dagger}\,(1\,-\, n_{i\overline{\sigma}})\, a_{j\sigma}\,+\,
\sum_{ij\sigma} {'}\, t_{ij}\,
a_{i\sigma}^{\dagger}\,(1\,-\, n_{i\overline{\sigma}})\, a_{j\sigma}\,
n_{j\overline{\sigma}}
$$
\begin{equation}
+\,\sum_{ij\sigma} {'}\, t_{ij}\,
a_{i\sigma}^{\dagger}\, n_{i\overline{\sigma}}\, a_{j\sigma}\,
(1\,-\, n_{j\overline{\sigma}})\,+\,
\sum_{ij\sigma} {'}\, t_{ij}\,
a_{i\sigma}^{\dagger}\, n_{i\overline{\sigma}}\, a_{j\sigma}\,
n_{j\overline{\sigma}}\,.
\label{eq:w7}
\end{equation}

But then, there is a problem: all the processes have the same coupling constants 
(are coming from the same term in the Hamiltonian, so singling out only the 
second and the third term as perturbation does not sound right. After a 
few-month deliberation I will not dwell upon, I have realized that the canonical 
perturbation in the version by Bogolyubov is the right method to follow. This
is because of the three following features of the method:
\begin{enumerate}
\item You can perform a perturbation in an invariant (operator) form (not only
on matrix elements),
\item If $H=H_{0}+H_{1}$ and $H_{1}$ is a perturbation part, then
in order to do the
calculations $H_{0}$ {\em does not have\/} to be diagonal, and you can select 
the perturbation term according to your wishes,
\item It bears a strong resemblance to the quantum mechanical perturbation 
approach when Hamiltonian splits into (large) diagonal blocks and (small) 
off-diagonal blocks, as I learnt earlier from the Stevens lectures [10].
\end{enumerate}

Formally, one can write the full Hamiltonian as operating in four subspaces

\begin{equation}
H_{0}\,=\, \sum_{ij\sigma} {'}\, t_{ij}\,
a_{i\sigma}^{\dagger}\, a_{j\sigma}\,=\, P_{0}\, H\, P_{0}\,+\,
P_{N}\, H\, P_{N}\,+\, P_{0}\, H\, P_{N}\,+\, P_{N}\, H\, P_{0}
\label{eq:w8}
\end{equation}
assuming (intuitively at this point) that $P_{0}+P_{N}={\bf 1}$, and that

\begin{equation}
P_{0}\, H\, P_{0}\,=\, \sum_{ij\sigma} {'}\, t_{ij}\,
a_{i\sigma}^{\dagger}\,(1\,-\, n_{i\overline{\sigma}})\, a_{j\sigma}\,
(1\,-\, n_{j\overline{\sigma}})\,,
\label{eq:w9}
\end{equation}

\begin{equation}
P_{N}\, H\, P_{N}\,=\, \sum_{ij\sigma} {'}\, t_{ij}\,
a_{i\sigma}^{\dagger}\, n_{i\overline{\sigma}}\, a_{j\sigma}\,
n_{j\overline{\sigma}}\,+\, U\, n_{i\uparrow}\, n_{i\downarrow}\,,
\label{eq:w10}
\end{equation}

\begin{equation}
P_{0}\, H\, P_{N}\,=\,
\sum_{ij\sigma} {'}\, t_{ij}\,
a_{i\sigma}^{\dagger}\, (1\,-\, n_{i\overline{\sigma}})\, a_{j\sigma}\,
n_{j\overline{\sigma}}\,,
\label{eq:w11}
\end{equation}

\begin{equation}
P_{N}\, H\, P_{0}\,=\,
\sum_{ij\sigma} {'}\, t_{ij}\,
a_{i\sigma}^{\dagger}\, n_{i\overline{\sigma}}\, a_{j\sigma}\,
(1\,-\, n_{j\overline{\sigma}})
\,=\, (P_{0}\, H\, P_{N})^{\dagger}\,.
\label{eq:w12}
\end{equation}

The explicit form of these projection operators {\em were not\/}
specified then and they need not to be. Understanding that took me 
some time.
Since every term contains process $\sim t_{ij}$, it is convenient 
to define the operator

\begin{equation}
H_{\epsilon}\,=\, H_{0}\,+\, \epsilon\, H_{1}\,.
\label{eq:w13}
\end{equation}

Obviously, $H\equiv H_{\epsilon =1}$ and

\begin{equation}
H_{0}\,=\, P_{0}\, H\, P_{0}
\,+\, P_{N}\, H\, P_{N}
\label{eq:w14}
\end{equation}

\begin{equation}
H_{1}\,=\, P_{0}\,H\, P_{N}\,+\, P_{N}\, H\, P_{0}\,.
\label{eq:w15}
\end{equation}

By introducing $\epsilon$ we will collect the terms of the same order
in $\epsilon$. In this manner, we will assume that $P_{0}HP_{0}\sim t$
and $P_{N}HP_{N}\sim U$ represent two different energy manifolds,
({\em the Hubbard subbands\/}) distant roughly by $U$, whereas 
$P_{0}HP_{N}$ represents rare hopping process of creating a doubly occupied 
site from two neighboring singly occupied sites $<i,j>$. So, even though 
all the hopping terms contain the same hopping matrix element $t_{ij}$,
$P_{0}HP_{0}$ will provide contribution $\sim t_{ij}$ while $P_{0}HP_{N}$ will
provide the contribution of $\sim t_{ij}/U$ in the first nontrivial order.

The last nontrivial feature is to select the unitary transformation matrix, with 
the help of which we remove part the hopping processes in the first order
and replace them by virtual processes exemplifying physical processes
in higher order. For that purpose, we have also proposed the following canonical 
transformation of the form

\begin{equation}
\tilde{H}_{\epsilon}\,=\,
e^{-i\,\epsilon\, S}\, H\, e^{i\,\epsilon\, S}\,\approx\,
H_{0}\,+\,\epsilon\, (H_{1}\,+\,i\, [H_{0}\,,\,S])\,-\,
\frac{1}{2} \,\epsilon^{2}\, \left(2i\, [H_{1}\,,\,S]\,-\,
[[H_{0}\,,\,S]\,,\,S]
\right)
\label{eq:w16}
\end{equation}
with $S=S^{\dagger}$.

The linear term $\sim \epsilon$ is absent when
\begin{equation}
H_{1}\,+\, i\,[\, H_{0},\, S\, ]\,=\, 0
\label{eq:w17}
\end{equation}
Under this condition
\begin{equation}
\tilde{H}_{\epsilon}\,=\,
H_{0}\,+\,\frac{1}{2} \, i\,\epsilon^{2}\, [\, H_{1},\,S\, ]
\label{eq:w18}
\end{equation}
Eq. (\ref{eq:w17}) has to be solved for $S$. The main problem is that
$H_{0}$ is not diagonal. So, it is more difficult to solve it the way
it was treated originally by Bogolyubov (cf. also the treatment
of electron-phonon case by Kittel). Thus, in order to proceed one tries
to solve (\ref{eq:w17}) projected onto the subspaces. Explicitly,
we can write

\begin{equation}
P_{0}\,H_{1}\, P_{N}\,+\, i \, P_{0}\, [H_{0},\, S]\, P_{N}\,=\, 0\,,
\label{eq:w19}
\end{equation}

\begin{equation}
P_{0}\,(H_{0}\, S\,-\, S\, H_{0})\, P_{0}\,=\, 
P_{N}\,(H_{0}\, S\,-\, S\, H_{0})\, P_{N}\,=\, 0\,.
\label{eq:w20}
\end{equation}
From the first equation we see that while $S$ must have nontrivial form
as it contains $H_{1}$, the next two equations are of trivial character.
This is because one can rewrite them in the form

\begin{equation}
(P_{0}\,H_{0}\, P_{0})\, (P_{0}\, S\, P_{0})\,-\,
(P_{0}\, S\, P_{0})\, P_{0}\,(H_{0}\, P_{0})\,=\, 0\,,
\label{eq:w21}
\end{equation}

\begin{equation}
(P_{N}\,H_{0}\, P_{N})\, (P_{N}\, S\, P_{N})\,-\,
(P_{N}\, S\, P_{N})\, (P_{N}\,H_{0}\, P_{N})\,=\, 0\,.
\label{eq:w22}
\end{equation}
In the original version we have taken a particular solution 
$P_{0,N}\, S\, P_{0,N}=\gamma_{0,N}P_{0,N}$. Equally good would
be to replace r.h.s. with an arbitrary function $f(P_{0,N})$
which has a Taylor expansion as $f(P_{0,N})=\alpha_{0,N}{\bf 1}+
\gamma_{0,N}P_{0,N}$ because of the property $ P_{0,N}^{2}= P_{0,N}$.

In effect, we are left with Eq. (\ref{eq:w17}), which projected
takes one of the two possible forms

\begin{equation}
P_{0}\, S\, P_{N}\,=\, [-i\, P_{0}\, H_{1}\, P_{N}\,+\,
(P_{0}\,H\, P_{0})\, (P_{0}\, S\, P_{N})]\,
(P_{N}\, H\, P_{N})^{-1}\,.
\label{eq:w23}
\end{equation}
Obviously this relation is well defined if the operator
$(P_{N}HP_{N})^{-1}$ exists, but we leave such problems to mathematical 
physicists. What is important for us instead is, that since the transformation
matrix appears on both l.h.s. and r.h.s. of (\ref{eq:w23}), we can try to solve
it by iterating the solution. In the zeroth order we assumed that on
r.h.s. of (\ref{eq:w23}) $(P_{0}S^{(0)}P_{N})=0$ and then

\begin{equation}
P_{0}\, S^{(1)}\, P_{N}\,=\, -i\, P_{0}\, H_{1}\, P_{N}\,/\,
P_{N}\, H\, P_{N}\,.
\label{eq:w24}
\end{equation}
The solution up to infinite order takes the form
\begin{equation}
P_{0}\, S\, P_{N}\,=\, -i\, P_{0}\, H_{1}\, P_{N}\,
(P_{N}\, H\, P_{N}\,-\,P_{0}\, H\, P_{0})^{-1}\,.
\label{eq:w25}
\end{equation}
The simplest approximation of the denominator is to replace it by 
an average $<P_{N}HP_{N}-P_{0}HP_{0}>\approx U$, where $U$ is the
energy difference between the centers of gravity of the Hubbard
subbands. In effect, after substituting (\ref{eq:w25}) to (\ref{eq:w18})
as well as its Hermitian conjugate $P_{N}SP_{0}$, we obtain the effective
Hamiltonian $\tilde{H}\equiv\tilde{H}_{\epsilon=1}$ up to $t^{2}/U$ as

\begin{equation}
\tilde{H}\,=\,
P_{0}\, \tilde{H}\, P_{0}\,+\, P_{N}\, H\, P_{N}\,,
\label{eq:w26}
\end{equation}
with

\begin{equation}
P_{0}\, \tilde{H}\, P_{0}\,=\, P_{0}\, H\, P_{0}\,-\,
P_{0}\, H\, P_{N}\, H\, P_{0}\, / \, U\,,
\label{eq:w27}
\end{equation}
and

\begin{equation}
P_{N}\, \tilde{H}\, P_{N}\,=\, P_{N}\, H\, P_{N}\,+\,
P_{N}\, H\, P_{0}\, H\, P_{N}\,.
\label{eq:w28}
\end{equation}
The first part of (\ref{eq:w27}) describes the dynamics of electrons in the
lower Hubbard subband for $n\neq 1$, whereas (\ref{eq:w28}) will
represent the same for $1<n\neq2$. In the following discussion
we limit ourselves to the situation with $n\neq 1$.

\subsection{Meaning of the t-J model}

Substituting (\ref{eq:w11}) and (\ref{eq:w12}) to (\ref{eq:w27}) and 
(\ref{eq:w28}), respectively we obtain explicitly [11]

$$
P_{0}\, \tilde{H}\, P_{0}\,=\, \sum_{ij\sigma} \, t_{ij}\,
a_{i\sigma}^{\dagger}\,(1\,-\, n_{i\overline{\sigma}})\, a_{j\sigma}\,
(1\,-\, n_{j\overline{\sigma}})
$$

$$
+\,\sum_{ij} {'}\, (2\, t_{ij}^{2}/U)\,
\left[ {\bf S}_{i}\,\cdot\,{\bf S}_{j}\,-\,\frac{1}{4}\,
\sum_{\sigma^{\prime}}\,
n_{i\sigma}\, (1\,-\,n_{i\overline{\sigma}})\, 
n_{j\sigma^{\prime}}\, (1\,-\, 
n_{j\overline{\sigma}^{\prime}})\right]
$$

$$
+\, \sum_{ijk}\, \frac{t_{ij}\,t_{jk}}{U}\,
\left[ a_{i\sigma}^{\dagger}\, (1\,-\, n_{i\overline{\sigma}})\, 
n_{j\overline{\sigma}}\,
(1\,-\, n_{j\sigma})\, a_{k\sigma}\,
(1\,-\, n_{k\overline{\sigma}})\right.
$$

\begin{equation}
\left. -\, a_{i\sigma}^{\dagger}\,
(1\,-\, n_{i\overline{\sigma}})\, {\bf S}_{j}^{-\sigma}\,
a_{k\overline{\sigma}}\, (1\,-\,n_{k\sigma})\right]\,,
\label{eq:w29}
\end{equation}
and

$$
P_{N}\, \tilde{H}\, P_{N}\,=\, \sum_{ij} \, t_{ij}\,
a_{i\sigma}^{\dagger}\, n_{i\overline{\sigma}}\, a_{j\sigma}\,
n_{j\overline{\sigma}}\,+\,
U\, \sum_{i}\, n_{i\uparrow}\, n_{i\downarrow}
$$

\begin{equation}
-\,\sum_{ij}\, \frac{2\, t_{ij}^{2}}{U}\,
a_{i\uparrow}^{\dagger}\, a_{i\downarrow}^{\dagger}
a_{j\downarrow}\, a_{j\uparrow}\,+\, \ldots\,.
\label{eq:w30}
\end{equation}

%The different terms in (\ref{eq:w29}) are represented schematically
%in Fig. 1. 
The first term in (\ref{eq:w29}) represents electron hopping from single-occupied
site onto an empty site $i$, the second term represents full
kinetic-exchange part in which, in general, the spin operators $\{
{\bf S}_{i}\}$ representing itinerant electrons are expressed in terms of
fermionic operators: ${\bf S}_{i}\equiv (S_{i}^{+},S_{i}^{-},S_{i}^{z})=
(a_{i\uparrow}^{\dagger} a_{i\downarrow}, a_{i\downarrow}^{\dagger}\, a_{i\uparrow}, (n_{i\uparrow}-n_{i\downarrow})/2)$. The third term
denotes hopping between three sites without and with spin flip, respectively,
in the middle $(j)$ site. One can easily see that in the limit of 
Mott-Hubbard insulator when the number of particles is conserved and
equal to unity {\em at each site\/} (i.e. $n_{i\uparrow}+n_{i\downarrow}=1$;
not only $<n_{i\uparrow}+n_{i\downarrow}>=1$!), then (\ref{eq:w29})
reduces properly to the Anderson (antiferromagnetic) kinetic exchange
Hamiltonian

\begin{equation}
P_{0}\, \tilde{H}\, P_{0}\,=\, 
\sum_{ij}\, \frac{2\, t_{ij}^{2}}{U}\,
\left({\bf S}_{i}\,\cdot\,{\bf S}_{j}\,-\,\frac{1}{4}\,\right)\,.
\label{eq:w31}
\end{equation}

In our original work [7,6] we have already noticed that the dynamics
requires introduction of projected fermionic creation, annihilation,
and particle-number operators

\begin{equation}
b_{i\sigma}^{\dagger}\,\equiv\, a_{i\sigma}^{\dagger}\,
(1\,-\, n_{i\overline{\sigma}})\,,
\label{eq:w32}
\end{equation}

\begin{equation}
b_{i\sigma}\,\equiv\, a_{i\sigma}\,
(1\,-\, n_{i\overline{\sigma}})\,,
\label{eq:w33}
\end{equation}

\begin{equation}
\nu_{i\sigma}\,\equiv\, b_{i\sigma}^{\dagger}\, b_{i\sigma}\,=\,
n_{i\sigma}\, (1\,-\, n_{i\overline{\sigma}})\,,
\label{eq:w34}
\end{equation}
which have {\em nonfermionic\/} anticommutation relations

\begin{equation}
\left\{ b_{i\sigma}\,,\, b_{j\sigma^{\prime}}^{\dagger}\right\}
\,=\,
\left[ (1\,-\, n_{i\overline{\sigma}})\,\delta_{\sigma\sigma^{\prime}}
\,-\, S_{i}^{-\sigma}\,
(1\,-\,\delta_{\sigma\sigma^{\prime}})\right]\,\delta_{ij}
\label{eq:w35}
\end{equation}

\begin{equation}
\left\{ b_{i\sigma}\,,\, b_{j\sigma^{\prime}}\right\}
\,=\,
\left\{ b_{i\sigma}^{\dagger}\,,\, b_{j\sigma^{\prime}}^{\dagger}\right\}
\,=\,0
\label{eq:w36}
\end{equation}

The property (\ref{eq:w35}) poses a problem as there is no reference 
occupation-number representation. Additionally, the hopping and
the exchange parts do not commute with each other, so the motion
of single electrons and the spin interactions (e.g. spin flip
processes) are entangled with each other and result in a {\em
strongly correlated (non-Fermi liquid)\/} metallic state.
What is equally important, the kinetic energy ($\sim t_{ij}$)
and the exchange ($t_{ij}^{2}/U$) can become comparable for a filling
close to the half filling and we can have a transition to the localized
state induced by the exchange interaction (see next Section) or to the
phase separation into antiferromagnetic insulating islands "floating"
in the ferromagnetic sea of $(1-n)$ holes [9]. This second state is
rather unstable as the Coulomb repulsion introduced by the holes 
i.e. by the positively charged ions of the background (usually
neglected in the calculations!) will destabilize the phase-separated
configuration.

Few comments are relevant at this point. First, the operators
(\ref{eq:w32})-(\ref{eq:w34}) were introduced by Hubbard [12]
under the name of atomic representation. The original Hubbard notation
is useful for multi-orbital models [13]. Nonetheless, it is used
sometimes also for the present model though, in our view, represents
an unnecessary formality. Second, it is sometimes important
to include the intersite Coulomb interaction, i.e. the term 

\begin{equation}
\frac{1}{2}\,\sum_{ij} {'}\, K_{ij}\,
n_{i}\, n_{j}\,.
\label{eq:w37}
\end{equation}

In that situation [9], the kinetic-exchange integral takes 
the form $J_{ij}=2t_{ij}^{2}/(U-K_{ij})$ and an additional
term $(1/2)\sum_{ij} {'} K_{ij}\,\nu_{i}\, \nu_{j}$ appears
in (\ref{eq:w29}). Finally, the essentially the same formalism
has been repeated later by other authors [14]. It is unfortunate
for us, those references are often quoted as those, from which
{\em the t-J model\/} (i.e. (\ref{eq:w29}), without 3-site
terms present) originated. We also comment on the effective t-J
model later on when we introduce the two-orbital model.

\subsection{Localization of holes in doped Mott insulator}

Intimately connected with the t-J model is the problem
of magnetic polarons [15] and the hole localization
induced by the kinetic exchange for the small number 
of correlated holes, $\delta=1-n\ll 1$. One can address the hole localization 
by starting from the metallic side. The ground state energy 
of (\ref{eq:w29}) can be estimated as follows 

\begin{equation}
E_{G}\,=\, \langle H\rangle\,=\,t\,
\sum_{<ij>\sigma}\, \left\langle \,
a_{i\sigma}^{\dagger}\,(1\,-\, n_{i\overline{\sigma}})\,
a_{j\sigma}\,(1\,-\, n_{j\overline{\sigma}})\,\right\rangle\,+\,
J\, \sum_{<ij>}\,\left\langle
{\bf S}_{i}\,\cdot\,{\bf S}_{j}\,-\,\frac{1}{4}\, \nu_{i}\, \nu_{j}\,
\right\rangle\,,
\label{eq:w38}  
\end{equation}
where we have included the terms involving only $z$ nearest 
neighbors $<i,j>$. In the spirit of Gutzwiller approach,
we can renormalize the hopping term by the band narrowing factor
$\Phi(\lambda)$, with $\lambda\equiv \langle \nu_{i}\nu_{j}/4-
{\bf S}_{i}\cdot{\bf S}_{j}\rangle/n^{2}$ and thus have

\begin{equation}
\frac{E_{G}}{N}\,=\, z\, t\, \Phi(\lambda)\,
\sum_{j(i)\sigma}\, \left\langle \,
a_{i\sigma}^{\dagger}\, a_{j\sigma}\,\right\rangle_{0}\,-\,
J\, z\, n^{2}\,\lambda
\label{eq:w39}
\end{equation}
where $\langle\ldots\rangle_{0}$ denotes the average for an
uncorrelated state.
Next, expand $\Phi(\lambda)=g_{0}+ g_{1}\lambda+ g_{2}\lambda^{2}$,
while $\langle a_{i\sigma}^{\dagger} a_{j\sigma}\rangle_{0}=
n_{\sigma}(1-n)$ expresses the hopping probability in the reference 
state without antiferromagnetic correlations $(\lambda=1/4)$.
For that value of $\lambda$ we have $g_{0}+ g_{1}/4 + g_{2}/16=1$.
In the opposite limit $(\lambda=1)$, we have a complete frozen
antiferromagnetic (N\`{e}el) state, without any hopping, i.e. 
$ g_{0}+ g_{1} + g_{2}=1$. Furthermore, writing the energy
in the form

\begin{equation}
\frac{E_{G}}{z\, N}\,=\, -\vert t\vert \, n\, (1\,-\, n)\,
(g_{0}\,+\, g_{1}\,\lambda\,+\, g_{2}\,\lambda^{2})\,-\,
J\, n^{2}\,\lambda\,,
\label{eq:w40}
\end{equation}
and minimizing it with respect to $\lambda$, we obtain the expression

\begin{equation}
4\, g_{2}\,\lambda\,=\,
-\frac{J\, n}{\vert t\vert\, (1\,-\, n)}\,+\, g_{1}\,.
\label{eq:w41}
\end{equation}

This equation must also fulfilled for $J=0$, when there are no 
antiferromagnetic correlations (i.e. when $\lambda=1/4$). This means that
from the above three conditions we can determine explicitly the coefficients
of the expansion of $\Phi(\lambda)$, which have the values:
$g_{0}=-1/9$, $g_{1}=8/9$, and $g_{2}=16/9$. Hence, the optimal value
of the variational parameter $\lambda$ is

\begin{equation}
\lambda\,=\,\frac{1}{32}\,
\frac{1\,+\, 9\, J}{\vert t\vert\, (1\,-\, n)}\,.
\label{eq:w42}
\end{equation}
This means that the full antiferromagnetic insulating state
$(\lambda=1)$ is achieved for

\begin{equation}
\delta\,=\,\delta_{C}\,=\,\left( 1\,+\,
\frac{8\,\vert t\vert}{3\, J}\right)^{-1}\,,
\label{eq:w43}
\end{equation}
and is of the order $\delta_{C}\simeq 0.1$ for $\vert t\vert/J=3$,
a reasonable value for La$_{2-x}$Sr$_{x}$CuO$_{4}$ system. 
The ground state energy is then

\begin{equation}
\frac{E_{G}}{z\, N}\,=\, -\vert t\vert \, n\, (1\,-\, n)\,
-\,\frac{1}{4}\, J\, n^{2}\,-\,\frac{9}{64}\,
\frac{J^{2}\, n^{3}}{\vert t\vert \,(1\,-\, n)}\,.
\label{eq:w44}
\end{equation}

This reasoning represents essentially the simplest mean-field-like
description of the transition from the strongly correlated 
paramagnetic metal to an antiferromagnetic insulator with frozen holes
at $\delta=\delta_{C}$.

\section{t-J model renewal: kinetic exchange as the source of real space 
pairing}

Until 1987, the t-J model was regarded as a generic model for explaining
the antiferromagnetism of Mott insulators such as the oxides NiO or CoO,
and relatively rarely considered for $n\neq 1$. Obviously, the real
oxides have both $3d$ and $2p$ orbital states as the valence-band states,
but it was argued strongly (and still is) [2,15,16], that the $p-d$
hybridization can be incorporated into an effective $d$-band Hubbard model,
although there were some exceptions to that idea [17]. The essential new idea,
about which I have learnt from the preprint of Ruckenstein et al. [18]
was to associate the kinetic exchange part with {\em real space pairing\/},
but there were quite few papers appearing at about the same time [19].
Soon, the {\em slave-boson\/} and {\em slave-fermion\/}
mean-field theories have been
formulated followed by a flood of papers, also about the stability of 
{\em $d$-wave superconducting
solution\/}. This is a fascinating story for itself, but I shall leave this to a 
separate occasion.

\subsection{t-J Hamiltonian as pairing Hamiltonian}

I would like to raise only one aspect of this story namely, the proper
formal expression of the pairing part [11]. As we have seen, we have
to express the whole Hamiltonian (\ref{eq:w29}) 
in terms of projected fermion operators
$b_{i\sigma}$ and $b_{i\sigma}^{\dagger}$. First, one notices that the 
spin operator has the same form in both fermion $\{ a_{i\sigma},
a_{i\sigma}^{\dagger}\}$ and in the projected $\{ b_{i\sigma},
b_{i\sigma}^{\dagger}\}$ representations. Explicitly, 

\begin{equation}
{\bf S}_{i}\,=\,\left( \, a_{i\uparrow}^{\dagger}\, a_{i\downarrow}\,,
\, a_{i\downarrow}^{\dagger}\, a_{i\uparrow}\,,\, 
(n_{i\uparrow}-n_{i\downarrow})/2\,\right)\,\equiv\,
\left( b_{i\uparrow}^{\dagger}\, b_{i\downarrow}\,,
\, b_{i\downarrow}^{\dagger}\, b_{i\uparrow}\,,\, 
(n_{i\uparrow}-n_{i\downarrow})/2\,\right)
\,.
\label{eq:w45}
\end{equation}
Therefore, the spin-singlet real-space paring operators should be
selected in the form $(i\neq j)$:

\begin{equation}
B_{ij}^{\dagger}\,\equiv\,\frac{1}{\sqrt{2}}\,
\left( b_{i\uparrow}^{\dagger}\, b_{j\downarrow}^{\dagger}\,-\,
b_{i\downarrow}^{\dagger}\, b_{j\uparrow}^{\dagger}\right)\,,
\label{eq:w46}
\end{equation}

\begin{equation}
B_{ij}\,\equiv\,-\frac{1}{\sqrt{2}}\,
\left( b_{i\uparrow}\, b_{j\downarrow}\,-\,
b_{i\downarrow}\, b_{j\uparrow}\right)
\,=\,
-\frac{1}{\sqrt{2}}\,
\left( a_{i\uparrow}\, a_{j\downarrow}\,-\,
a_{i\downarrow}\, a_{j\uparrow}\right) \,(1\,-\, n_{i\uparrow})\,
(1\,-\, n_{j\downarrow})\,.
\label{eq:w47}
\end{equation}
In this representation, we have the following identity

\begin{equation}
B_{ij}^{\dagger}\, B_{ij}\,\equiv\,
-\left( {\bf S}_{i}\,\cdot\,{\bf S}_{j}\,-\,
\frac{1}{4}\, \sum_{\sigma\sigma^{\prime}}\,
\nu_{i\sigma}\, \nu_{j\sigma^{\prime}}\right)\,.
\label{eq:w48}
\end{equation}
One should note at this point that the representation of 
(\ref{eq:w48}) through the operators $B_{ij}^{\dagger}$ and
$B_{ij}$ is completely equivalent to that through the spins
and projected particle-number operators. In other words,
the spin ordering and the nearest-neighbor pairing should be treated
on the same footing.

Substituting (\ref{eq:w48}) to $P_{0}\tilde{H}P_{0}$ we obtain
the effective t-J Hamiltonian in the following closed form
when the 3-site terms are also included [11]

\begin{equation}
\tilde{H}\,\equiv\, P_{0}\tilde{H}P_{0}\,=\,
\sum_{ij\sigma}\, t_{ij}\, 
b_{i\sigma}^{\dagger}\, b_{j\sigma}\,-\,
\sum_{ijk}\, 2\, t_{ij}\, t_{jk}/U\,
B_{ij}^{\dagger}\, B_{jk}\,.
\label{eq:w49}
\end{equation}
Obviously, it is assumed, that $t_{ij}\neq 0$ only for $i\neq j$.
Furthermore, the pairing part vanishes by itself if $i=j$ as 
$B_{ii}^{\dagger}=0$. This means that the pairing order parameter
will have a nontrivial dependence in both real and reciprocal
(${\bf k}$) spaces.

Transforming the Hamiltonian (\ref{eq:w49}) to ${\bf k}$ space, we
obtain pairing Hamiltonian, which takes the following form when there are
phase differences between the neighbors from the left and the right sides.

\begin{equation}
P_{0}\tilde{H}P_{0}\,=\,
\sum_{{\bf k}\sigma}\, \epsilon_{{\bf k}}\, 
b_{{\bf k}\sigma}^{\dagger}\, b_{{\bf k}\sigma}\,-\,
\frac{2\, t^{2}}{U}\, \sum_{{\bf k}{\bf k}^{\prime}}\, 
\gamma_{{\bf k}}\, \gamma_{{\bf k}^{\prime}}\,
B_{{\bf k}^{\prime}{\bf -k}}^{\dagger}\, 
B_{{\bf -k}^{\prime}{\bf k}}\,,
\label{eq:w50}
\end{equation}
where
\begin{equation}
B_{{\bf k}^{\prime}{\bf -k}}^{\dagger}\,\equiv\,
\frac{1}{\sqrt{2}}\,
\left( b_{{\bf k}\uparrow}^{\dagger}\, b_{{\bf -k}\downarrow}\,-\,
b_{{\bf k}\downarrow}^{\dagger}\, b_{{\bf -k}\uparrow}\right)
\label{eq:w51}
\end{equation}
and

\begin{equation}
b_{{\bf k}\sigma}^{\dagger}\,=\,\frac{1}{\sqrt{N}}\,
\sum_{{\bf R}_{i}} \, e^{-i\,{\bf k}\,\cdot\,{\bf R}_{i}}
a_{i\sigma}^{\dagger}\, (1\,-\, n_{i\overline{\sigma}})
\label{eq:w52}
\end{equation}
From the form (\ref{eq:w52}) we see that also the operators 
$ b_{{\bf k}\sigma}^{\dagger}$ (and $ b_{{\bf k}\sigma}$)
have a nontrivial many-particle character. The only simple 
property here is the separability of the pairing potential, i.e.
$V_{{\bf k}{\bf k}^{\prime}}=-(2t^{2}/U)\gamma_{{\bf k}}
\gamma_{{\bf k}^{\prime}}\equiv V_{{\bf k}}V_{{\bf k}^{\prime}}$,
where $\gamma_{{\bf k}}$ depends on the solution symmetry (for example,
for an extended s-wave solution for the square lattice
$\gamma_{{\bf k}}=\epsilon_{{\bf k}}/t$; see also below).

The solution of t-J model in the forms (\ref{eq:w49}) or (\ref{eq:w50})
is the subject of an intense discussion, but this absolutely
fundamental aspect of the model will not be touched upon here.
We hope to return to this aspect of the exchange mediated superconductivity
elsewhere.

\subsection{Three remarks on the pairing Hamiltonian with 3-site terms}

First remark concerns the formal form of $\gamma_{{\bf k}}$. Namely,
when taking the Fourier transform of the part $\sim t_{ij}t_{jk}$
in (\ref{eq:w49}) and assuming that for given neighbor (e.g. in
the square lattice) the operator $B_{ij}$ for $j$ to the right
from $i$ acquires the phase $\varphi_{x}$ with respect to that to 
the left. In effect, the part $\gamma_{{\bf k}}$ of the pairing potential
can be expressed as

\begin{equation}
\gamma_{{\bf k}}\,=\,
e^{i\,k_{x}\, a\,+\,\varphi_{x}}\,+\,
e^{-i\,k_{x}\, a}\,+\,
e^{i\,k_{y}\, a\,+\,\varphi_{y}}\,+\,
e^{-i\,k_{y}\, a}\,,
\label{eq:w53}
\end{equation}
or equivalently 

$$
\gamma_{{\bf k}}\,=\,
2\, \cos\left( k_{x}\,+\,\frac{\varphi_{x}}{2}\right)\, \cos
\left(\frac{\varphi_{x}}{2}\right)\,+\,
2\, i\, \sin\left( k_{x}\,+\,\frac{\varphi_{x}}{2}\right)\, \sin
\left(\frac{\varphi_{x}}{2}\right)
$$
\begin{equation}
+\,
2\, \cos\left( k_{y}\,+\,\frac{\varphi_{y}}{2}\right)\, \cos
\left(\frac{\varphi_{y}}{2}\right)\,+\,
2\, i\, \sin\left( k_{y}\,+\,\frac{\varphi_{y}}{2}\right)\, \sin
\left(\frac{\varphi_{y}}{2}\right)
\label{eq:w54}
\end{equation}
This form represents the most general ${\bf k}$ dependence of the
pairing when the 3-site terms are included.

Second remark concerns the equivalence (\ref{eq:w48}) of spin
degrees of freedom and the real-space pairing operators (note however,
that r.h.s is nonzero for $i=j$). Therefore, when decoupling r.h.s
of (\ref{eq:w49}) we should take into account a possibility of have nonzero
both averages. In the Hartree-Fock-like approximation this decoupling 
for $k=i$ should take the form

\begin{equation}
B_{ij}^{\dagger}\, B_{ij}\simeq\, 
\langle B_{ij}^{\dagger}\rangle\, B_{ij}\,+\,
\langle B_{ij}\rangle\, B_{ij}^{\dagger}\,-\,
\langle B_{ij}^{\dagger}\rangle\, \langle B_{ij}\rangle\,
\,-\,\left[ \, S_{i}^{z}\, \langle S_{j}^{z}\rangle\, \,+\,
\langle S_{j}^{z}\rangle\, S_{i}^{z}\,-\,
\langle S_{i}^{z}\rangle\, \langle S_{j}^{z}\rangle\,\right]\,+\,\ldots
\label{eq:w55}
\end{equation}

So, in general, the coexistence of antiferromagnetism (or spin density wave)
and the paired state is possible and should be studied carefully.

Last remark concerns mapping of the projected Hamiltonian onto an effective
fermionic Hamiltonian, as proposed [11,20] on early stage of the theory 
development. Namely, introducing the concept of band narrowing 
$\Phi=\Phi(\lambda)$ one can have that

$$
\tilde{H}\,=\, \Phi\, \sum_{ij\sigma}\, t_{ij}\, a_{i\sigma}^{\dagger}\,
a_{j\sigma}\,-\, (1\,-\,\Phi)\,\sum_{ij}\, (2t_{ij}^{2}/U)\,
B_{ij}^{\dagger}\, B_{ij}
$$
\begin{equation}
-\,\Phi\, (1\,-\,\Phi) \,\sum_{ijk}\, (2t_{ij}t_{kj}/U)
B_{ij}^{\dagger}\, B_{kj}\,,
\label{eq:w56}
\end{equation}
where $0\leq \Phi\leq 1$ is the band narrowing factor. Macroscopically 
$\Phi$ can be regarded as a degree of itineracy of electrons. Likewise,
$(1-\Phi)$ is the "fraction" of localized electrons. In (\ref{eq:w56})
the two-site and three-site terms have different renormalization factors.
Hence, is tempting to take, as earlier, $\langle B_{ij}^{\dagger} B_{ij}\rangle$
as the variational parameter $\lambda$. Then, as before $\Phi=\Phi(\lambda)$.
It is also tempting then to associate the pairing only with the 3-site part,

but as we are aware of, this possibility has not been studied
explicitly as yet. In such model the limit of the antiferromagnetic insulator
is reached naturally for $\delta\rightarrow 0$ and the pseudogap 
appearance is associated with short-range antiferromagnetic correlations.

\section{Kondo interaction and real-space hybrid pairing
as an extension of the t-J model concepts}

\subsection{Deep and shallow impurity cases}

Shortly after deriving the effective Hamiltonian (\ref{eq:w49}) and determining 
the principal features of antiferromagnetic phase I have turned my attention
to the Wolff model of magnetic impurity [21] and the corresponding
generalization of the canonical transformation [7,6],
as it has been proposed
and discussed by us earlier [22]. The Wolff model differs from the original 
Anderson-impurity model [23] only slightly, as it treats both the impurity 
electron state and the band states of a metal, the impurity is immersed in,
in the tight binding approximation. Explicitly, the Hamiltonian
of the system is proposed in the form

$$
H\,=\, \sum_{ij\sigma} {'} {'}\, t_{ij}\, a_{i\sigma}^{\dagger}\,
a_{j\sigma}\,+\, \epsilon_{d}\,(n_{0\uparrow}\,+\, n_{0\downarrow})\,+\,
U\, n_{0\uparrow}\, n_{0\downarrow}
$$
\begin{equation}
+\,(1\,+\,\gamma)\,\sum_{j\sigma}\, t_{0j}\,\left(
a_{0\sigma}^{\dagger}\, a_{j\sigma}\,+\,
a_{j\sigma}^{\dagger}\, a_{0\sigma}\right)\,.
\label{eq:w57}
\end{equation}

Here we assume that the impurity is at the 0th site, with $\epsilon_{d}$
as the energy for the one-electron impurity state. The intraatomic Coulomb
energy for the two electron impurity state is $U$ (and zero for the 
metallic host sites $j\neq 0$). The hopping strength between the impurity
and the host is specified by $\gamma$. The double primed summation
in (\ref{eq:w57}) excludes the terms with either $i=0$ or $j=0$;
$\epsilon_{d}$ is the impurity energy-level position with respect to that for band states, which is taken as $\epsilon_{jj}=0$.

We were interested in the strong correlation limit when $\vert (1+\gamma)
t_{0j}\vert$ is much smaller than $U$. What was (and still is) new
in our original approach is, that we have considered separately the 
{\em shallow-\/} and {\em deep-impurity limits\/}. This turned out to have
a direct relation respectively to the Kondo and to the mixed-valence (or 
heavy-fermion) limits of correlated electrons. Also, it has relevance 
to the concept of $p-d$ pairing in the high-temperature-superconductivity era 
[24].
As these reminiscences are probably too long-winded already, I provide
only the essence of the argument leading to the pairing. But first,
summarize the results for the impurity situation.

In the interesting us case of a shallow impurity we have the impurity
(correlated) state immersed in the Fermi sea of the metal. In that situation
singly occupied impurity state is relatively close to the metal Fermi energy
$\epsilon_{F}$ but its doubly occupied state is high in energy and therefore,
can be realized via virtual excitations (hence, transformed out again). In 
effect, the effective Hamiltonian in the second order takes rather complicated 
form

$$
\tilde{H}\,\equiv\,P_{0}\,\tilde{H}\, P_{0}\,=\, 
\sum_{ij\sigma} {'} {'}\, t_{ij}\, a_{i\sigma}^{\dagger}\,
a_{j\sigma}\,+\, (1\,+\,\gamma)\,\sum_{j\sigma}\, t_{0j}\,
(1\,-\, n_{0\overline{\sigma}})\,
\left( a_{0\sigma}^{\dagger}\, a_{j\sigma}\,+\,
a_{j\sigma}^{\dagger}\, a_{0\sigma}\right)
$$
\begin{equation}
+\,\epsilon_{d}\,(\nu_{0\uparrow}\,+\, \nu_{0\downarrow})\,+\,
\frac{2\,(1\,+\,\gamma)^{2}}{U\,+\,\epsilon_{d}}\,
\sum_{j}\, t_{0j}^{2}\,
\left( {\bf S}_{0}\,\cdot\, {\bf S}_{j}\,-\,\frac{1}{4}\,
\sum_{\sigma\sigma^{\prime}}\, \nu_{0\sigma}\, n_{j\sigma^{\prime}} \right)
\,.
\label{eq:w58}
\end{equation}

This effective interaction contains both the antiferromagnetic impurity-host
kinetic exchange interaction (the last term, which can be called the 
{\em non-local Kondo interaction\/}) and the residual hopping between the 
impurity and the metallic host. This form, at first look, seems to be 
of limited use, as {\em there is\/} still part of the hopping left between the 
host and the impurity. As we show in the following, this type of situation
is specific for the heavy fermions and high-T$_{C}$ systems,
since the valence $(p)$ states play an active role in the electric
conductivity.

For the sake of completeness, we provide also the effective Hamiltonian 
in the {\em deep-impurity\/} case, i.e. when the $\epsilon_{d}$ level
is deep below the Fermi level. It is

$$
\tilde{H}\,=\, 
\sum_{ij\sigma} {'} {'}\, t_{ij}\, a_{i\sigma}^{\dagger}\,
a_{j\sigma}\,+\, 
\epsilon_{d}\,(\nu_{0\uparrow}\,+\, \nu_{0\downarrow})\,-\,
\frac{(1\,+\,\gamma)^{2}}{\epsilon_{d}}\,\sum_{j\sigma}\, t_{0j}^{2}\,
n_{j\sigma}
$$
\begin{equation}
-\,\frac{2\, (1\,+\,\gamma)^{2}\, U}{\epsilon_{d}\,(U\,+\,\epsilon_{d})}\,
\left[
\sum_{j}\, t_{0j}^{2}\,
\left( {\bf S}_{0}\,\cdot\, {\bf S}_{j}\,-\,\frac{1}{4}\,
\sum_{\sigma}\, n_{j\sigma} \right)\,+\,\frac{1}{2}\,\sum_{j}\,
t_{i0}\, t_{j0}\, S_{0}^{\sigma}\, a_{i\overline{\sigma}}^{\dagger}\,
a_{j\sigma}\right]
\,.
\label{eq:w59}
\end{equation}

This Hamiltonian has a direct correspondence to that derived by
Schrieffer and Wolff [25] for the Anderson-impurity model.
The last term represents the Kondo interaction, with three-site
spin-flip term included (here we have explicitly assumed, that
in the present limit $n_{0\uparrow}+n_{0\downarrow}=1$, i.e. we have
localized moment at the impurity (0th) site).

\subsection{Hybrid pairing in two-orbital system}

After the t-J model had been interpreted in the categories of electron
pairing, a natural question arose if one can do the same for the
hybridized (two-orbital) system such as the Anderson-lattice model.
This idea has been put into writing subsequently [24] by starting
from the Hamiltonian

$$
H\,=\, 
\sum_{mn\sigma} {'}\, t_{mn}\, c_{m\sigma}^{\dagger}\,
c_{n\sigma}\,+\, 
\epsilon_{f}\,\sum_{i\sigma}\, N_{i\sigma}\,+\, U
\sum_{i}\, N_{i\uparrow}\,N_{i\downarrow}
$$
\begin{equation}
+\,\sum_{im}\, V_{im}\,
\left( a_{i\sigma}^{\dagger}\, c_{n\sigma}\,+\,
c_{m\sigma}^{\dagger}\, a_{i\sigma}\right)\,,
\label{eq:w60}
\end{equation}
where the $(i,j)$ label correlated atomic ($a=d$ or $f$) states,
$(m,n)$ label delocalized $c$ states, and $N_{i\sigma}=
a_{i\sigma}^{\dagger}a_{i\sigma}$. $V_{im}$ represents hybridization 
integral. As in the case of shallow impurity, we assume that $\vert V_{im}
\vert\ll U$, but $\vert \epsilon_{f}\vert\sim\vert V_{im}\vert$, so
{\em not the whole\/} hybridization term can be transformed out. Now, decompose
the hybridization term as before, i.e. 

\begin{equation}
a_{i\sigma}^{\dagger}\, c_{m\sigma}\,+\, 
c_{m\sigma}^{\dagger}\, a_{i\sigma}\,=\,
(1\,-\, N_{i\overline{\sigma}})\,
(a_{i\sigma}^{\dagger}\, c_{m\sigma}\,+\,
c_{m\sigma}^{\dagger}\, a_{i\sigma})\,+\,
N_{i\sigma}\, 
(a_{i\sigma}^{\dagger}\, c_{m\sigma}\,+\,
c_{m\sigma}^{\dagger}\, a_{i\sigma})
\,.
\label{eq:w61}
\end{equation}
The first term on r.h.s. represents the interstate ($a-c$) hopping processes
which do not involve $U$ (double occupancies of atomic states), while
the second does involve $U$ and hence leads to higher-order mixing processes.
The basic idea introduced at this point [11] is to canonically transform out the
second term only and to replace it by an effective interaction incorporating
higher-order virtual processes. Leaving the details [24], the effective 
Hamiltonian with real-space pairing has the form

$$
\tilde{H}\,=\, 
\sum_{mn\sigma} {'}\, t_{mn}\, c_{m\sigma}^{\dagger}\,
c_{n\sigma}\,+\, 
\epsilon_{f}\,\sum_{i\sigma}\, N_{i\sigma}
\,(1\,-\, N_{i\overline{\sigma}})\,+\,
\sum_{im\sigma}\, V_{im}\,
(1\,-\, N_{i\overline{\sigma}})\, 
(a_{i\sigma}^{\dagger}\, c_{m\sigma}\,+\, 
c_{m\sigma}^{\dagger}\, a_{i\sigma})
$$
\begin{equation}
-\,\sum_{imn}\, \frac{2\, V_{mi}\, V_{in}}{U\,+\,\epsilon_{f}}\,
B_{im}^{\dagger}\, B_{in}\,,
\label{eq:w62}
\end{equation}
where the hybrid-pairing operators are defined as

\begin{equation}
B_{im}^{\dagger}\,\equiv\,\frac{1}{\sqrt{2}}\,
\left( a_{i\uparrow}^{\dagger}\, c_{m\downarrow}^{\dagger}\,-\, 
a_{i\downarrow}^{\dagger}\, c_{m\uparrow}^{\dagger}\right)\,
\left( 1\,-\, N_{i\overline{\sigma}}\right)\,.
\label{eq:w63}
\end{equation}

This model can be transformed again to the effective Fermi-liquid
form and the proper form is:

$$
\tilde{H}\,=\, 
\sum_{{\bf k}\sigma}\, \epsilon_{{\bf k}}\, n_{{\bf k}\sigma}\,
\,+\, 
\tilde{\epsilon}_{f}\,\sum_{i\sigma}\, N_{i\sigma}\,+\,
\sum_{{\bf k}\sigma}\, \tilde{V}_{{\bf k}\sigma}\,
\left( a_{{\bf k}\sigma}^{\dagger}\, c_{{\bf k}\sigma}\,+\, 
c_{{\bf k}\sigma}^{\dagger}\, a_{{\bf k}\sigma}\right)
$$
\begin{equation}
-\,\frac{2}{U\,+\,\epsilon_{f}}\,
\sum_{{\bf kq}}\, V_{{\bf k}}\, V_{{\bf q}}\,
B_{{\bf k},{\bf -k}}^{\dagger}\, B_{{\bf q},{\bf -q}}
\,,
\label{eq:w64}
\end{equation}
where the tilted quantities are renormalized by correlation, e.g.

\begin{equation}
\tilde{V}_{{\bf k}}\,\equiv\, q_{\sigma}\, V_{{\bf k}}\,=\,
q_{\sigma}\, \sum_{j(m)}\, e^{i\,{\bf k}\,\cdot\,{\bf R}_{j}}\, V_{jm}\,,
\label{eq:w65}
\end{equation}
and $q_{\sigma}=(1-n_{f})/(1-n_{f\sigma})$, with $n_{f\sigma}=
\langle a_{i\sigma}^{\dagger}a_{i\sigma}\rangle$. Also, the factor
$(1\,-\, N_{i\overline{\sigma}})$ is absent in $B_{im}^{\dagger}$
(cf. Eq. (\ref{eq:w63})). This Hamiltonian can be solved in BCS 
approximation, but the results will not be reproduced here. Likewise, 
we will omit here the application of (\ref{eq:w62}) to the description of
high-T$_{C}$ superconductivity with hybrid $p-d$ pairing, as it would
most probably require the inclusion of the 4th order
($d-d$ or $f-f$) interactions.

\subsection{Kinetic exchange in orbitally degenerate systems and possibility
of spin-triplet pairing}

This topic grew into an independent discipline of its own after publication
of the papers by Kugel and Khomskii [26], Lacroix and Cryot [27], and
Inagaki [28]. Some aspects of the topic are reviewed in this issue 
by Ole\'s [29]. We have also derived the kinetic-exchange Hamiltonian
for a partial filing of the band [30], as well as have applied it to explain 
ferromagnetism of CoS$_{2}$ [31]. In general, inclusion of the orbital
degeneracy allows for a natural explanation of the appearance of {\em 
ferromagnetic
Mott-Hubbard insulators\/} in conjunction with the {\em orbital ordering\/}
of antiferromagnetic type in systems such as K$_{2}$CuF$_{4}$.

In connection with this, a question has arisen, whether the ferromagnetic
kinetic exchange can produce spin-triplet pairing. This question is particularly 
important because it has been shown earlier, that the Hund's rule coupling
can lead not only to ferromagnetism, but also to the superconducting pairing
[32]. Such an effective model has been proposed and analyzed in detail
in the strong-correlation limit very recently [33]. These last results
will be submitted for a publication shortly.

\section{Concluding remarks}

In this overview we have concentrated on the idea of kinetic exchange
as it was derived 30 years ago and its subsequent application to the systems
with real space pairing mediated by this exchange interaction. By no means
it is a complete survey. For example, we have ignored all the subsequent formal
development of the model (see e.g. [34]). Also, we have disregarded the effect
of electron-lattice coupling on the effective t-J model with pairing [35].
Nonetheless, what I hope I have sketched here is the analytical structure 
of the t-J model in various strongly correlated systems, in which magnetism
and superconductivity seem to have a common origin, although it has not as yet
been proved conclusively, that the kinetic exchange is the origin of both
of them. Future will show.

\subsection*{Acknowledgment}

In addition to my colleagues and students mentioned in the Foreword
to this volume, I would like to thank Dr. Robert Podsiad\l{}y
for his technical help. This work was supported by 
Foundation of Polish Science
(FNP) and the
Grant No. 1 P03B 001 29 from Ministry of Science and Higher Education.
Part of the work mentioned is elaborated under the auspices of COST P-16
Network of European Science Foundation.

\end{document}